# Analyzing the Low Power Wireless Links for Wireless Sensor Networks

Md. Mainul Islam Mamun, Tarek Hasan-Al-Mahmud, Sumon Kumar Debnath, Md. Zahidul Islam

**ABSTRACT**— There is now an increased understanding of the need for realistic link layer models in the wireless sensor networks. In this paper, we have used mathematical techniques from communication theory to model and analyze low power wireless links. Our work provides theoretical models for the link layer showing how Packet Reception Rate vary with Signal to Noise Ratio and distance for different modulation schemes and a comparison between MICA2 and TinyNode in terms of PRR.

## 1 INTRODUCTION

Wireless sensor network protocols are often evaluated through simulations that make simplifying assumptions about the link layer, such as the binary perfect-reception-within-range model. Several recent empirical studies [1] [2] [3] have questioned the validity of these assumptions. These studies have revealed the existence of three distinct reception regions in a wireless link: connected, transitional, and disconnected. The transitional region is often quite significant in size, and is generally characterized by high-variance in reception rates and asymmetric connectivity. Particularly, in dense deployments such as those envisioned for sensor networks, a large number of the links in the network (even higher than 50%) can be unreliable because of the transitional region. Because of its inherent unreliability and extent, the transitional region can have a considerable impact on the performance of higher layer protocols. For instance, in [1] it is shown that the dynamics of even the simplest flooding mechanism and the topology of data gathering trees constructed in dense sensor networks can be significantly affected by the existence of such unreliable links.

————————————————

● *Md. Mainul Islam Mamun with the Computer Science and Telecommunication Engineering Department,, Noakhali Science and Technology University, Noakhali.*
● *Tarek Hasan-Al-Mahmud with the Computer Science and Telecommunication Engineering Department, Noakhali Science and Technology University, Noakhali.*
● *Sumon Kumar Debnath with the Computer Science and Telecommunication Engineering Department,, Noakhali Science and Technology University, Noakhali.*
● *Md. Zahidul Islam, PhD Fellow, Chonnam National University, South Korea, On leave from the Information and Communication Engineering Department, Islamic University, Kushtia.*

In order to address this need, some recent works [3] [7] [8] have proposed new link models based on empirical data. Woo et al. present a curve fit model for the link packet reception rate as a function of distance for a specific radio and a specific environment. Zhou et al. present the radio irregularity model (RIM), which provides for anisotropic radio coverage. Cerpa et al. have undertaken a large collection of empirical radio measurements using the SCALE tool and have used it to generate statistical models for different specific radio and environmental settings.

In this study, we make use of these analytical tools, particularly expressions for the log-normal shadowing path loss model (to model the environment) and the bit error performance of various modulation and encoding schemes with respect to the signal to noise ratio (to model the radio). We combine these to derive expressions for the packet reception rate as a function of distance for different settings. We use these in turn to determine the width of the transitional region, allowing us to quantify and analyze how it is impacted by important radio parameters such as modulation.

## 2 BACKGROUND

### 2.1 The Wireless Channel

When an electromagnetic signal propagates, it may be diffracted, reflected and scattered. These effects have two important consequences on the signal strength. First, the signal strength decays exponentially with respect to distance. Second, for a given distance d, the signal strength is random and log-normally distributed about the mean distance-dependent value.

Due to the unique characteristics of each environment, most radio propagation models use a combination of analytical and empirical methods. One of the most common radio propagation models, for large and small [12] coverage systems, is the log-normal shadowing path loss model [13]

$$PL(d) = PL(d_0) + 10n\log_{10}(d/d_0) + X_\sigma \qquad (1)$$

Where d is the transmitter-receiver distance, $d_0$ is a reference distance, n is the path loss exponent (rate at which signal decays), and $X\sigma$ is a zero-mean Gaussian random variable (in dB) with standard deviation σ (shadowing effects) The received signal strength *(Pr)* at a distance d is the output power of the transmitter minus *PL(d)*.

## 2.2 The Radio

To facilitate the explanation of the radio model, this subsection assumes NRZ encoding. The steps followed to derive the radio model are similar to the ones in [9]. Let $P_i$ be a Bernoulli random variable, where $P_i$ is 1 if the packet is received and 0 otherwise. Then, for *r* transmissions, the packet reception rate is defined by

$$\frac{1}{r}\sum_{i=1}^{r} P_i$$

Since $P_i$'s are i.i.d. random variables, by the weak law of large numbers PRR can be approximated by *E[Pi]*, where E[Pi] is the probability of successfully receiving a packet. When NRZ is used, 1 Baud = 1 bit, hence the probability *p* of successfully receiving a packet is:

$$P = (1 - P_e)^{8l}(1 - P_e)^{8(f-l)} = (1 - P_e)^{8f} \qquad (2)$$

Where f is the frame size, l is the preamble (both in bytes). And $P_e$ is the probability of bit error. $P_e$ depends on the modulation scheme.

Given a transmitting power $P_t$, the SNR at a distance *d* is (all powers in dB):

$$\gamma(d) = P_t - PL(d) - P_n \qquad (3)$$

Henceforth, the PRR at a distance d for the NRZ encoding is:

$$p(d) = (1 - P_e)^{8f} \qquad (4)$$

With the aim of obtaining the radius of the different regions, let us bound the connected region to PRRs greater than 0.9, and the transitional region to values between 0.9 and 0.1.

## 2.3 Different Modulation Schemes and PRR

The different Packet Reception Rates (PRR), *p(d)* according to the NRZ encoding scheme, frame, *f* and preamble length, *l* is shown in TABLE 1.

**TABLE 1**: Modulation Schemes and PRR

| Modulation Techniques | Packet Reception Rate (PRR) |
|---|---|
| NCFSK | $p(d) = (1 - \frac{1}{2} exp^{-\gamma(d)/2})^{8f}$ |
| CFSK | $p(d) = (1 - Q(\sqrt{\gamma(d)}))^{8f}$ |
| BPSK | $p(d) = (1 - Q(\sqrt{2\gamma(d)}))^{8f}$ |
| DPSK | $p(d) = (1 - \frac{1}{2} exp^{-\gamma(d)})^{8f}$ |

## 2.4 MICA2 and TinyNode

The MICA2 is 3G mote module used for enabling bw-power wireless, sensor networks. It supports the wireless remote reprogramming. The TinyNode 584 is an ultra-low power OEM module that provides a simple and reliable way to add wireless communication to sensors, actuators, and controllers.

## 3 SIMULATIONS OF THE MODELS

Fig. 1 shows Receiver Response for CFSK and NCFSK modulation schemes for NRZ radio when *f=50 bytes*. From this curve we found that for The highest PRR, the SNR values are 11 *dBm* and 20 *dBm* for CFSK and NCFSK respectively.

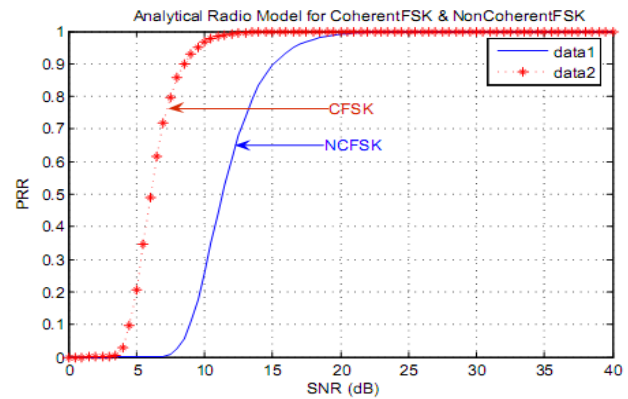

Fig. 1 Receiver Response: CFSK and NCFSK, NRZ radio, *f*=50 *bytes*

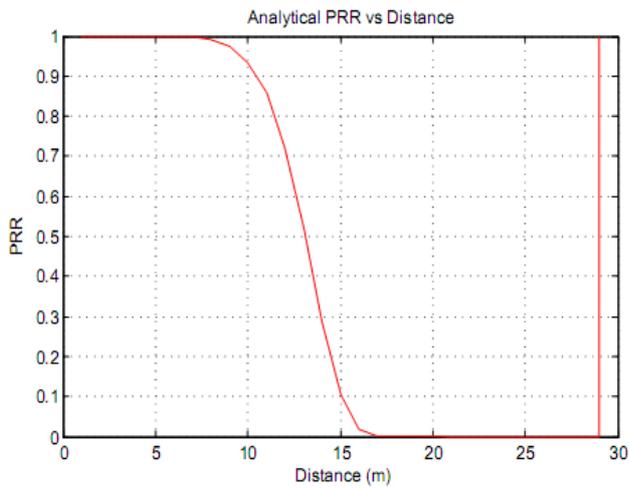

Fig. 2 Analytical PRR vs Distance: NCFSK

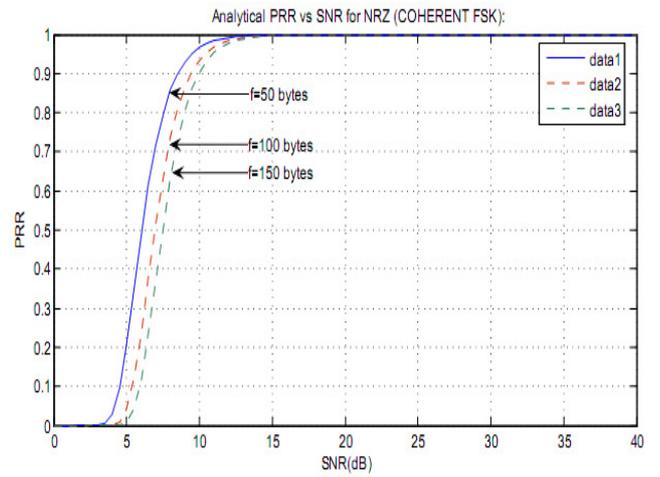

Fig. 5 Receiver Response for Different Frames Size for Coherent FSK

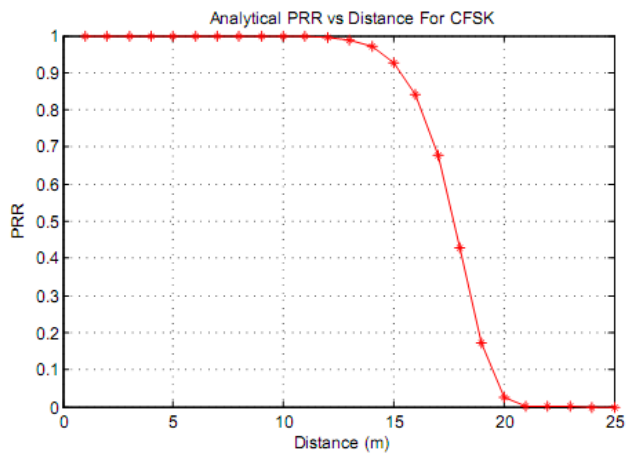

Fig. 3 Analytical PRR vs Distance: CFSK

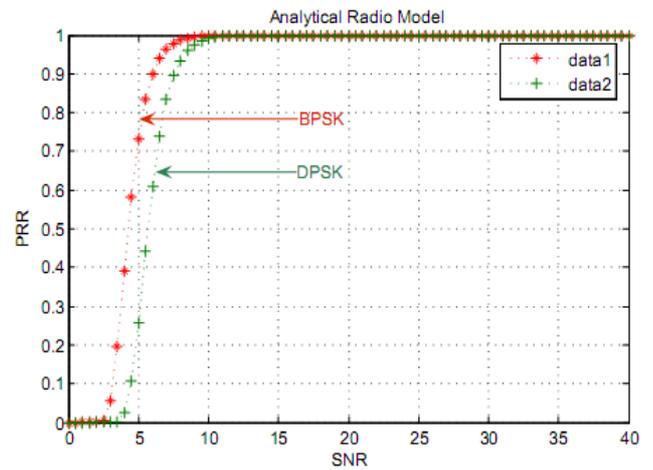

Fig. 6 Receiver Response: BPSK and DPSK, NRZ radio, *f*=50 *bytes*

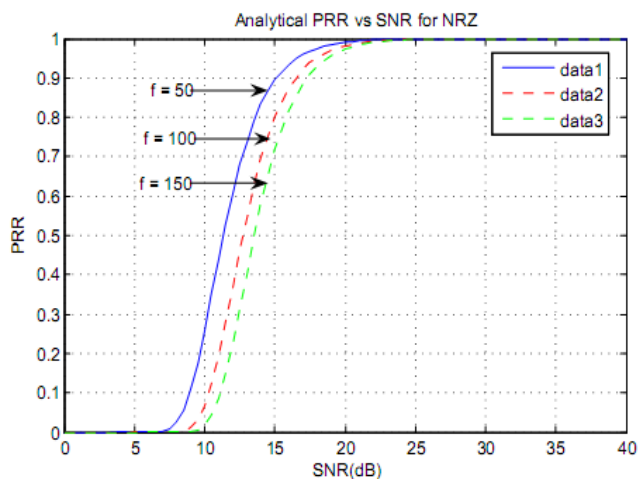

Fig. 4 Receiver Response for Different Frames Size for Non-Coherent FSK

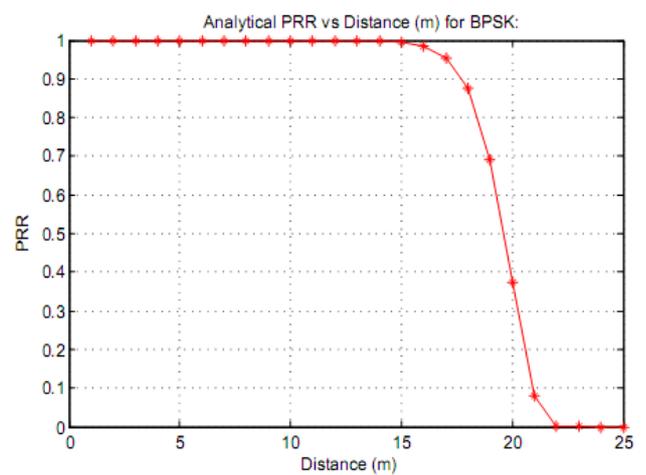

Fig. 7 Analytical PRR vs Distance: BPSK

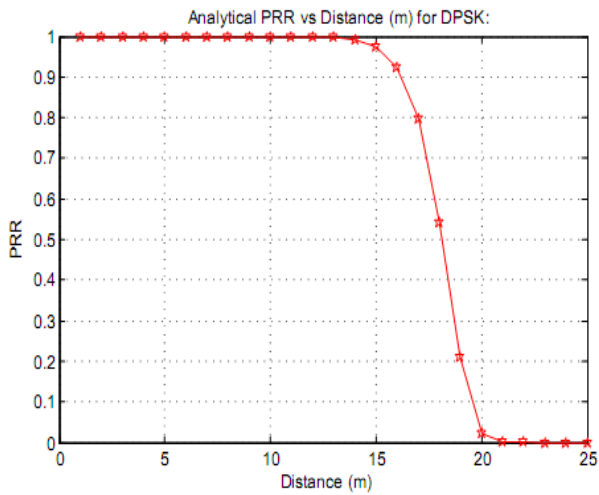

Fig. 8 Analytical PRR vs Distance: DPSK

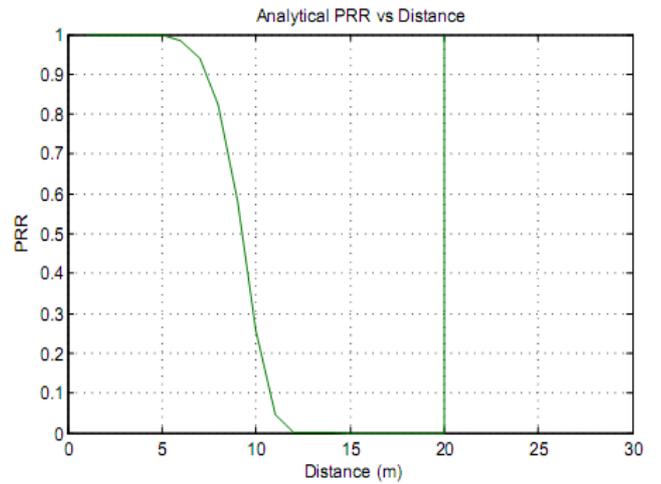

Fig. 11 Analytical PRR vs Distance: MICA2
for $P_t = +5$ *dBm*, $P_n = -104$ *dBm*

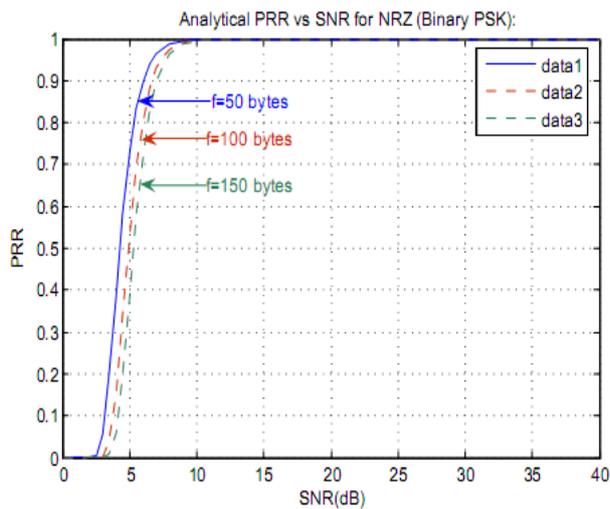

Fig. 9 Receiver Response for Different Frames Size for Binary Phase Shift Keying

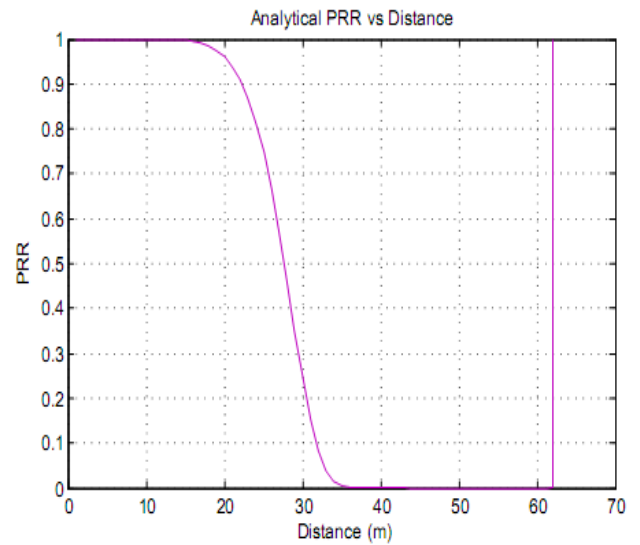

Fig. 12 Analytical PRR vs Distance: TinyNode
for $P_t = +12$ *dBm*, $P_n = -116$ *dBm*

Fig. 2 shows Analytical PRR vs Distance for NCFSK. The PRR is maximum up to 7.5 *m*, the Transitional Region starts from 11 m and ends at 29 *m*. The connected region is from 0 *m* to 11 *m*. The connected region is from 0 *m* to 15 *m* for CFSK, which is shown in Fig. 3. Fig. 4 and Fig. 5 show the Receiver Response for Different Frames Size for Non-Coherent FSK and Coherent FSK. Fig. 6 shows the Receiver Response for both the BPSK and DPSK modulation and NRZ encoding for frame size,*f*=50 *bytes*. The PRR is maximum up to the SNR value 7.5 *dBm* for BPSK and 11 *dBm* for DPSK. Fig. 7 and Fig. 8 show the PRR vs Distance curve. The PRR value is maximum from 0 to 15 *m* for BPSK and from 0 to 13 *m* for DPSK. By observing the figures 7, 8, we can say that the TR is from 17.5 *m* for BPSK and 16.4 *m* for DPSK. Fig. 9 and 10 show the Receiver Response for Different Frames for BPSK and DPSK respectively. The PRR vs distance curves for MICA2 and TinyNode are shown in Fig. 11 and 12 respectively. The length of connected region is 7.5 *m* for MICA2 and 22 *m* for TinyNode. The TR for MICA2 is from 7.5 *m* to 20 *m*. For TinyNode this TR is from 21 *m* to 61 *m*.

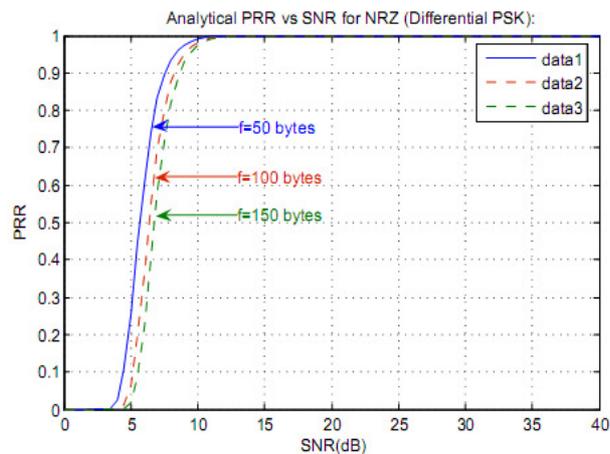

Fig. 10 Receiver Response for Different Frames Size for Differential Phase Shift Keying

## 4 CONCLUSIONS

We have presented the PRR variations with SNR and distances for Non-Coherent Frequency Shift Keying, Coherent Frequency Shift Keying, Binary Phase Shift Keying and Differential Phase Shift Keying for NRZ encoding scheme with MICA2 node. We have found that the CFSK is better than NCFSK in term of PRR, because in order to get the maximum PRR the required SNR value is less for CFSK than NCFSK. The distance with maximum for CFSK PRR is greater than NCFSK. The Connected Region for CFSK and NCFSK is 15 *m* and 11 *m* respectively. For any encoding scheme, as the frame size increases, the SNR bounds increase (curves shift right) which leads to smaller Connected Region. The BPSK performs better than DPSK in term of PRR. Also, the connected region is larger for BPSK than DPSK. This CR for BPSK is similar to CFSK. This analysis also yields that the TinyNode performs better than MICA2 in term of PRR and Connected Region. The methodology presented here can be easily extended to other radios that use different modulation and encoding scheme.

## REFERENCES

[1] D. Ganesan, B. Krishnamachari, A. Woo, D. Culler, D. Estrin and S. Wicker. "Complex Behavior at Scale: An Experimental Study of Low-Power Wireless Sensor Networks". UCLA CS Technical Report UCLA/CSD-TR 02-0013, 2002.

[2] Jerry Zhao, Ramesh Govindan, "Understanding packet delivery performance in dense wireless sensor networks", Proceedings of the 1st international conference on Embedded networked sensor systems, November 05-07, 2003, Los Angeles, California, USA

[3] Alec Woo, Terence Tong , David Culler, "Taming the underlying challenges of reliable multihop routing in sensor networks", Proceedings of the 1st international conference on Embedded networked sensor systems, November 05-07, 2003, Los Angeles, California, USA

[4] G. Zhou, T. He, S. Krishnamurthy, and J. A. Stankovic. "Impact of radio irregularity on wireless sensor networks". International Conference on Mobile Systems, Applications and Services, 2004.

[5] A. Cerpa, J. L. Wong, L. Kuang, M. Potkonjak and D. Estrin. "Statistical Model of Lossy Links in Wireless Sensor Networks". CENS Technical Report 0041, April 2004.

[6] S. Y. Seidel and T. S. Rapport. "914 MHz Path Loss Prediction Model for Indoor Wireless Communication in Multi oored Building". In IEEE Transactions on Antennas and Propagation, volume 40(2), pages 207-217, February 1992.

[7]Theodore S. Rappapport. "Wireless Communications: Principles and Practice". Prentice Hall.

[8] D. Lal, A. Manjeshwar, F. Herrmann, E. Uysal-Biyikoglu, and A. Keshavarzian. "Measurement and characterization of link quality metrics in energy constrained wireless sensor networks". In IEEE GLOBECOM, pages 172--187, December 2003.